\documentclass[aip,cha,amsmath,amssymb,reprint,author-year,author-numerical]{revtex4-1}
\usepackage{epsfig}
\usepackage{color}
\usepackage{graphicx}
\usepackage{dcolumn}
\usepackage{bm}
\input epsf

\usepackage[normalem]{ulem} 

\graphicspath{{figures/}}

\begin{document}


\title{Optimal synchronization of directed complex networks}

\author{Per Sebastian Skardal}
\email{persebastian.skardal@trincoll.edu}
\affiliation{Department of Mathematics, Trinity College, Hartford, CT 06106, USA}

\author{Dane Taylor}
\affiliation{Carolina Center for Interdisciplinary Applied Mathematics, Department of Mathematics, University of North Carolina, Chapel Hill, NC 27599, USA}

\author{Jie Sun}
\affiliation{Department of Mathematics, Clarkson University, Potsdam, NY 13699, USA}

\begin{abstract}
We study optimal synchronization of networks of coupled phase oscillators. We extend previous theory for optimizing the synchronization properties of undirected networks to the important case of directed networks. We derive a generalized synchrony alignment function that encodes the interplay between network structure and the oscillators' natural frequencies and serves as an objective measure for the network's degree of synchronization. Using the generalized synchrony alignment function, we show that a network's synchronization properties can be systematically optimized. This framework also allows us to study the properties of synchrony-optimized networks, and in particular, investigate the role of directed network properties such as nodal in- and out-degrees. For instance, we find that in optimally rewired networks the heterogeneity of the in-degree distribution roughly matches the heterogeneity of the natural frequency distribution, but no such relationship emerges for out-degrees. We also observe that a network's synchronization properties are promoted by a strong correlation between the nodal in-degrees and the natural frequencies of oscillators, whereas the relationship between the nodal out-degrees and the natural frequencies has comparatively little effect. This result is supported by our theory, which indicates that synchronization is promoted by a strong alignment of the natural frequencies with the left singular vectors corresponding to the largest singular values of the Laplacian matrix.
\end{abstract}


\maketitle

\begin{quotation}
Synchronization is vital to the functionality of many natural and engineered systems~\cite{Strogatz2003,Pikovsky2003,Arenas2008PR}, including cardiac pacemaker cells~\cite{Glass1988}, circadian rhythms~\cite{Yamaguchi2003Science}, Josephson junction arrays~\cite{Wiesenfeld1996PRL}, and power grids~\cite{Motter2013NaturePhysics}. This has generated considerable interest in optimizing the synchronization properties of networks~\cite{Brede2008PLA,Buzna2009PRE,Kelly2011Chaos,Scafuti2015PRE,Pinto2015PRE,Skardal2014PRL}. In a recent publication, we developed a theoretical framework for optimizing the synchronization properties of undirected networks of heterogeneous oscillators~\cite{Skardal2014PRL}. Here, we extend this theory to the important case of directed networks and derive a generalized synchrony alignment function (SAF) that can be used to systematically optimize a network's synchronization properties. Furthermore, this approach allows us to examine which structural properties promote synchronization in directed networks. Potential applications include systems where strong synchronization is essential for efficient functionality, including cardiac electrophysiology~\cite{KroghMadsen2012,Karma2013Rev}, synthetic cell engineering~\cite{Prindle2012Nature}, and power grid dynamics~\cite{Rohden2012PRL,Dorfler2013PNAS,Skardal2015SciAdv,Witthaut2016PRL}.
\end{quotation}


\section{Introduction}\label{sec1}
The tendency for large groups of individual units to reach consensus despite having heterogeneous dynamical properties has served as strong motivation for scientists to study synchronization of coupled dynamical systems~\cite{Strogatz2003,Pikovsky2003}. A paradigmatic model for studying synchronization and the emergence of collective behavior was developed by Kuramoto~\cite{Kuramoto1984}, who showed that under appropriate conditions, the dynamics of $N$ oscillators can be reduced to the evolution of $N$ phases, $\theta_i$, for $i=1,\dots,N$. When placed on a network that indicates the oscillators' interaction, the phases evolve according to
\begin{align}
\dot{\theta}_i=\omega_i+K\sum_{j=1}^NA_{ij}H_{ij}(\theta_j-\theta_i),\label{eq:Kuramoto}
\end{align}
where $\omega_i$ is the natural frequency of oscillator $i$, $K$ is the global coupling strength, $H_{ij}(\theta)$ is a $2\pi$-periodic coupling function, and $A$ is the adjacency matrix that encodes the network structure such that $A_{ij}=1$ if a link exists from node $j$ to node $i$. In many cases the topology of the network is assumed to be undirected, so that $A=A^T$; however, here we will consider the more general case of a possibly directed network topology~\cite{Restrepo2005Chaos}. Furthermore, to ensure for the possibility of synchronization from Eq.~(\ref{eq:Kuramoto}), we assume that the coupling frustration\cite{Skardal2015PREb} is sufficiently small, $|H_{ij}(0)/\sqrt{2}H_{ij}'(0)|\ll1$.

Extensive research has demonstrated that the interplay between dynamics and network structure has nonlinear effects on the synchronization of a network. For instance, different networks can give rise to different synchronization patterns~\cite{Restrepo2005PRE,GomezGardenes2007PRL,GomezGardenes2011PRL,Skardal2012PRE,Restrepo2014EPL,Skardal2015PRE}, and at the same time synchronization can be utilized to analyze the properties of a given network~\cite{Arenas2006PRL,Skardal2015}. The macroscopic synchronization dynamics of Eq.~(\ref{eq:Kuramoto}) is typically quantified using the classical Kuramoto order parameter $r$ defined by the complex number
\begin{align}
re^{i\psi}=\frac{1}{N}\sum_{j=1}^Ne^{i\theta_j},\label{eq:OrderParameter}
\end{align}
which represents the centroid of phases  $\{\theta_i\}$ after mapping them onto the complex unit circle. In particular, $r$ ranges between $0$ and $1$, representing completely incoherent and perfectly synchronized states, respectively, with intermediate values representing partially synchronized states. In a recent publication~\cite{Skardal2014PRL}, we developed a theoretical framework for optimizing the synchronization properties of a given undirected network, as defined by maximizing the order parameter $r$. In particular, we derived the {\it synchrony alignment function} (SAF), a functional that encodes the interplay between local dynamical properties (i.e., the oscillators' natural frequencies $\{\omega_i\}$) and the network structure (via the eigenvalues and eigenvectors of the network Laplacian matrix). We showed that the SAF can be used to systematically optimize the synchronization properties of a network under a wide variety of constraints. 

As in the study of other network-coupled dynamical processes, extending analysis for undirected networks to directed networks represents a non-trivial hurdle for theoretical and practical progress\cite{Sanchez2002PRL,Restrepo2008PRL,Lentz2012PRE}. In the case of identical oscillators, our understanding of synchronization on directed networks is well-developed through the framework of Master Stability Functions~\cite{Pecora1998PRL}, and in fact, directed coupling can be utilized to achieve improved and moreover optimized synchronization, although sometimes at the expense of reduced robustness~\cite{Nishikawa2006PRE,Nishikawa2010PNAS,Ravoori2011PRL}.

In this paper, we extend our previous results~\cite{Skardal2014PRL} for optimizing synchronization of heterogeneous oscillators in undirected networks to the case of directed networks. We derive a generalized SAF and demonstrate its utility with several examples. In particular, we show that the generalized SAF can be used optimize synchronization under several constraints: (i)~choosing the oscillators' natural frequencies for a given network; (ii)~arranging a set of pre-chosen natural frequencies on a given network; and (iii)~building a network for a given set of natural frequencies. We emphasize that our approach allows for efficient optimization of a network's synchronization properties based on objective measures, not heuristics. Furthermore, the generalized SAF approach allows us to investigate the dynamical and topological properties of synchrony-optimized networks. For example, we study the directed network properties of nodal in- and out-degrees and find that synchronization is promoted by a strong, positive correlation between the nodal in-degrees and magnitude of the natural frequencies; however, the relationship between the nodal out-degrees and natural frequencies has comparatively little effect on synchronization. We additionally observe that synchronization is enhanced by a negative correlation between the natural frequencies of neighboring oscillators. These results extend previous research~\cite{Skardal2014PRL,Brede2008PLA,Buzna2009PRE,Kelly2011Chaos,Scafuti2015PRE,Pinto2015PRE} that observed similar correlations to promote synchronization, but only considered the case of undirected networks.

The remainder of this paper is organized as follows: In Sec.~\ref{sec2}, we present our theoretical framework and derive the generalized SAF. In Sec.~\ref{sec3}, we provide several examples of optimizing synchronization using the generalized SAF. In Sec.~\ref{sec4} we study the effects that directed network properties have on optimal synchronization and the effect that optimizing synchronization has on directed network structure. In Sec.~\ref{sec5}, we study the relationship between structural and dynamical properties of synchrony-optimized networks. In Sec.~\ref{sec6}, we conclude with a summary and discussion of our results.

\section{Derivation of the generalized synchrony alignment function (SAF)}\label{sec2}
The derivation of the generalized SAF begins as in Ref.~\cite{Skardal2014PRL}, by considering the dynamics of Eq.~(\ref{eq:Kuramoto}) in a state of strong synchronization, i.e., $r\approx1$. For a generic network, such a state can be obtained in several ways, typically by either (i) sufficiently increasing the coupling strength $K$ or (ii) sufficiently decreasing the spread or standard deviation of the natural frequencies. In fact, up to a rescaling of time, these two actions are equivalent. In this regime, the oscillators become strongly clustered about the mean phase $\psi$ such that $|\theta_j-\theta_i|\ll1$ for all $1\le i,j\le N$, and thus Eq.~(\ref{eq:Kuramoto}) can be linearized to
\begin{align}
\dot{\theta}_i\approx\tilde{\omega}_i-KH'(0)\sum_{j=1}^NL_{ij}\theta_j,\label{eq:Theory01}
\end{align}
where we have assumed for simplicity that each coupling function is the same, i.e., $H_{ij}(\theta)=H(\theta)$ for all $1\le i,j\le N$, we have defined an effective natural frequency
\begin{align}
\tilde{\omega}_i=\omega_i+KH(0)k_i^{\text{in}} , \label{eq:Theory02}
\end{align}
$L$ is the Laplacian matrix defined for directed networks with entries defined
\begin{align}
L_{ij} = \delta_{ij}k_i^{\text{in}}-A_{ij},\label{eq:Theory03}
\end{align}
where $\delta_{ij}$ is the Kronecker delta, and the nodal in- and out-degrees are given by
\begin{align}
k_i^{\text{in}}=\sum_{j=1}^NA_{ij},\hskip4ex k_i^{\text{out}}=\sum_{j=1}^NA_{ji}.\label{eq:Theory04}
\end{align}
In vector form, Eq.~(\ref{eq:Theory01}) can be more conveniently rewritten as
\begin{align}
\dot{\bm{\theta}}=\bm{\tilde{\omega}}-KH'(0)L\bm{\theta},\label{eq:Theory05}
\end{align}
where $\bm{\theta}=[\theta_1,\dots,\theta_N]^T$ and $\bm{\tilde{\omega}}=[\tilde{\omega}_1,\dots,\tilde{\omega}_N]^T$.

We now search for a phase-locked solution of Eq.~(\ref{eq:Theory05}), $\dot{\bm{\theta}}=\Omega\bm{1}$, where $\Omega$ represents the collective frequency of the networks. To find $\bm{\omega}$ and $\Omega$, we require the Moore-Penrose pseudoinverse $L^\dagger$ that satisfies $LL^\dagger L=L$ and $L^\dagger LL^\dagger=L^\dagger$~\cite{BenIsrael1974}. In the case of directed networks where $L^T\ne L$, the formulation for the pseudoinverse requires the singular value decomposition of $L$,
\begin{align}
L=U\Sigma V^T,\label{eq:Theory06}
\end{align}
which is defined by the set of $2N$ equations
\begin{align}
L\bm{v}^i=\sigma_i\bm{u}^i,\hskip4ex L^T\bm{u}^i=\sigma_i\bm{v}^i.\label{eq:Theory07}
\end{align}
In particular, the left and right singular vectors $\bm{u}^i$ and $\bm{v}^i$ populate the columns of $U$ and $V$, respectively, and the singular values $\sigma_i$ populate the diagonal matrix $\Sigma=\text{diag}(\sigma_1,\dots,\sigma_N)$. The singular vectors are normalized such that each set $\{\bm{u}^i\}_{i=1}^N$ and $\{\bm{v}^i\}_{i=1}^N$ forms an orthonormal basis for $\mathbb{R}^N$ and $U$ and $V$ are orthogonal matrices. Importantly, the singular values $\sigma_i$ are all real and non-negative~\cite{Golub}. Since each row of $L$ sums to zero, the first singular value $\sigma^1$ is precisely zero and corresponds to right singular vector $\bm{v}^1=[1,\dots,1]^T$. If the network is strongly connected, i.e., any node can be reached from any other node, then all other singular values are positive and can be ordered $0=\sigma_1<\sigma_2\le\dots\le\sigma_N$. Finally, defining $\Sigma^\dagger=\text{diag}(0,\sigma_2^{-1},\dots,\sigma_N^{-1})$, the pseudoinverse $L^\dagger$ is given by 
\begin{align}
L^\dagger=V\Sigma^\dagger U^T=\sum_{j=2}^N\frac{\bm{v}^j\bm{u}^{jT}}{\sigma_j}.\label{eq:Theory08}
\end{align}

The formulation of the pseudoinverse $L^\dagger$ in terms of the singular value decomposition in the case of a directed network has a number of implications on the collective dynamics of networks. First, the collective frequency $\Omega$ of the synchronized population is not necessarily equal to the mean $\langle\tilde{\omega}\rangle$, but rather is given by a weighted average of the entries of $\bm{\tilde{\omega}}$~\cite{Skardal2015}. Specifically, the weights are given by the entries of the first left singular vector $\bm{u}^1$, 
\begin{align}
\Omega=\frac{\langle\bm{u}^1,\bm{\tilde{\omega}}\rangle}{\langle\bm{u}^1,\bm{1}\rangle}=\frac{\sum_iu_i^1\tilde{\omega}_i}{\sum_iu_i^1},\label{eq:Theory09}
\end{align}
where $\langle \bm{x},\bm{y}\rangle=\bm{x}^T\bm{y}=\sum_ix_iy_i$ denotes the inner product. Returning to the dynamics of Eq.~(\ref{eq:Theory05}), we enter the rotating frame $\theta_i\mapsto\theta_i+\Omega t$ and apply the pseudoinverse $L^\dagger$ to find the steady-state solution
\begin{align}
\bm{\theta}^*=\frac{L^\dagger\bm{\tilde{\omega}}}{KH'(0)}.\label{eq:Theory10}
\end{align}
To evaluate the order parameter $r$ for a state given by Eq.~(\ref{eq:Theory10}), we note that with a suitable shift in initial conditions the mean phase $\psi$ of the population can be set to zero. Thus, the order parameter can be expanded to
\begin{align}
r\approx 1-\frac{\|\bm{\theta}^*\|^2}{2N},\label{eq:Theory11}
\end{align}
where $\|\cdot\|$ denotes the Euclidean norm. Finally, noting that the squared norm can be evaluated by $\| x\|^2=\langle x,x\rangle$, we arrive at
\begin{align}
r\approx1-\frac{J(\bm{\tilde{\omega}},L)}{2K^2H'^2(0)},\label{eq:Theory12}
\end{align}
where $J(\cdot,\cdot)$ is the {\it generalized synchrony alignment function} defined as
\begin{align}
J(\bm{\tilde{\omega}},L)=\frac{1}{N}\sum_{j=2}^N\frac{\langle\bm{u}^j,\bm{\tilde{\omega}}\rangle^2}{\sigma_j^2}.\label{eq:Theory13}
\end{align}

Equations~(\ref{eq:Theory12}) and (\ref{eq:Theory13}) serve as an objective measure of the degree of synchronization for a given network. Importantly, Eq.~(\ref{eq:Theory12}) implies that maximizing the order parameter $r$ is equivalent to minimizing the generalized SAF, thus providing a framework for optimizing the synchronization properties of a directed network. Specifically, the SAF encodes the interplay between the local oscillator dynamics $\bm{\tilde{\omega}}$ and the network structure, here represented by the left singular vectors $\bm{u}^j$ and singular values $\sigma_j$, and can be used to optimize synchronization analytically and/or algorithmically, depending on the given constraints. In the next section, we will demonstrate the utility of the generalized SAF in optimizing synchronization in directed networks for several classes of constraints.

Before demonstrating the utility of the generalized SAF via concrete examples, a few remarks are in order. The SAF $J(\bm{\tilde{\omega}},L)$ for a directed network given by Eq.~(\ref{eq:Theory13}) represents the generalization from the undirected case. The analysis presented here for directed networks include undirected networks as a special case: For undirected networks where $L^T=L$ the singular values are precisely the eigenvalues of $L$, and the left and right singular vectors become equal and correspond to the eigenvectors of $L$. In the directed case, the left singular vectors and singular values play equivalent roles as the eigenvectors and eigenvalues in the undirected case; they determine the alignment of the local oscillator dynamics $\bm{\tilde{\omega}}$ with the network structure. We note that the effective natural frequencies $\bm{\tilde{\omega}}$ depend not only on the natural frequencies $\bm{\omega}$, but also on the coupling strength $K$, nodal in-degree $\bm{k}^{\text{in}}$, and coupling function via $H(0)$. Therefore, optimization of synchronization using the SAF can generally depend on the coupling strength. However, if the oscillator coupling is not frustrated, i.e., $H(0)=0$, then the system simplifies, $\bm{\tilde{\omega}}=\bm{\omega}$, and optimization via the SAF is independent of coupling strength. 



\begin{figure*}[t]
\centering
\epsfig{file =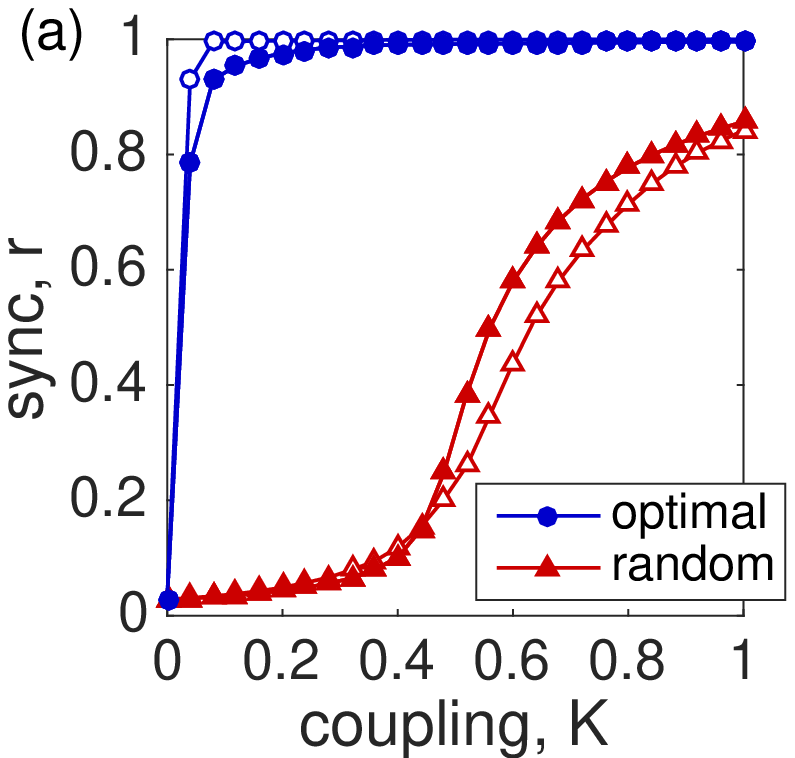, clip =,width=0.25\linewidth }
\epsfig{file =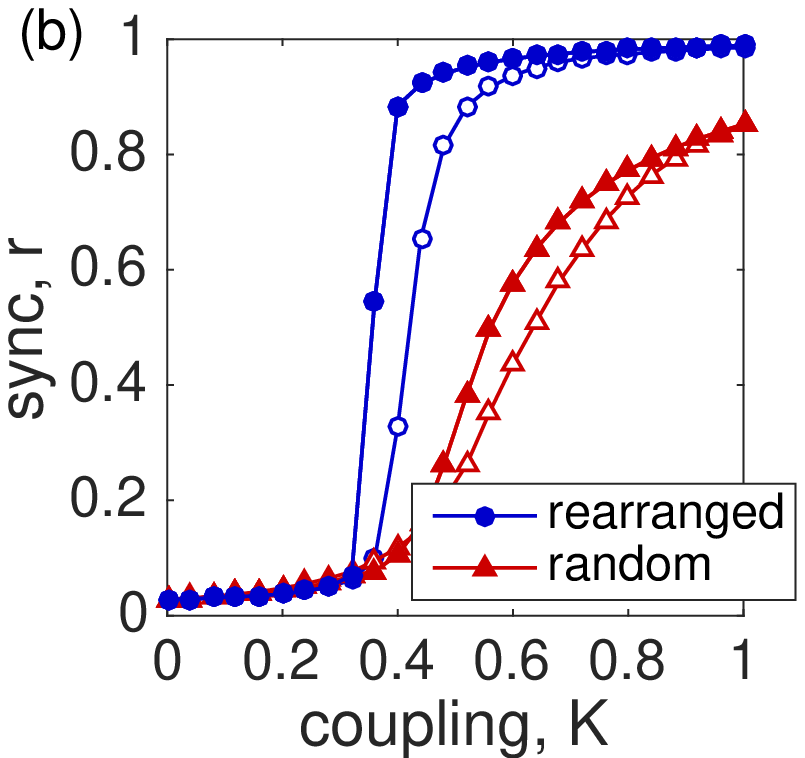, clip =,width=0.25\linewidth }
\epsfig{file =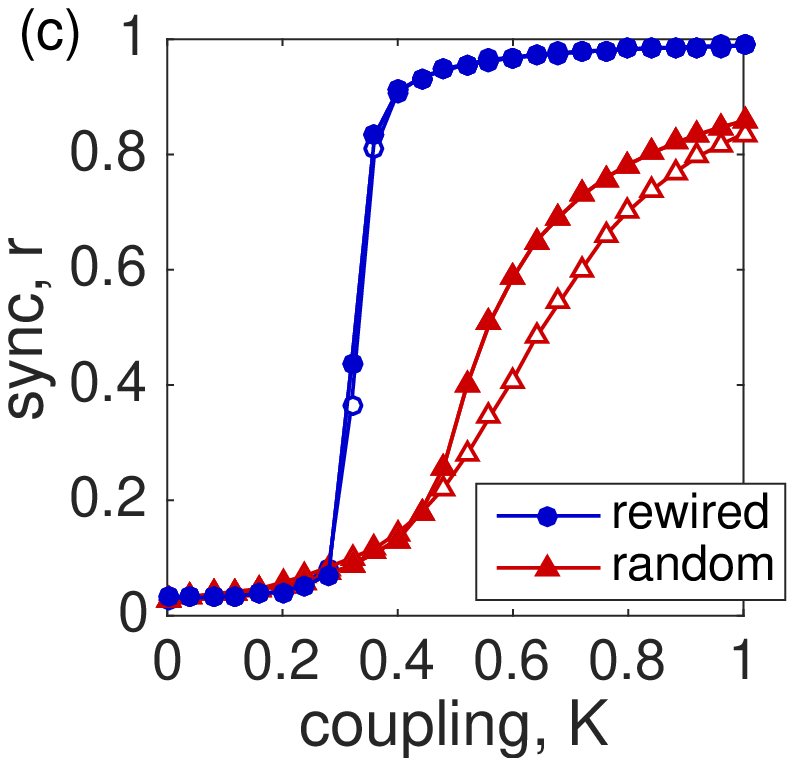, clip =,width=0.25\linewidth }
\caption{(Color online) Optimal synchronization of directed networks: Synchronization profiles $r$ vs $K$ for (a) oscillator allocation, (b) oscillator arrangement, and (c) network construction for normally distributed natural frequencies. Erd\H{o}s-R\'{e}nyi (ER) and scale-free (SF) networks of size $N=1000$ in (a) and (b) and $N=500$ in (c) with mean degree $\langle k\rangle=4$ and $\gamma=3$ are considered. Results represent an average over 50 network realizations of each network type. Results for the optimal and random networks are plotted in blue circles and red triangles, respectively, and ER and SF networks are indicated with filled and unfilled markers, respectively.}\label{fig1}
\end{figure*}

\section{Optimizing network synchronization}\label{sec3}
We now demonstrate with several examples the utility of the generalized SAF for optimizing the synchronization properties of directed networks. We will study three classes of constrained optimization problems: (i)~allocation of natural frequencies given a fixed network (Sec.~\ref{sec3sub1}); (ii)~arrangement of a set of pre-chosen natural frequencies given a fixed network (Sec.~\ref{sec3sub2}); and (iii)~construction of a network for a set of pre-chosen natural frequencies (Sec.~\ref{sec3sub3}). In all cases, we will consider natural frequency vectors with mean zero, $\langle\omega\rangle=\frac{1}{N}\sum_{j=1}^N\omega_i=0$, and fixed, nonzero standard deviation, $\sigma^2=\frac{1}{N}\sum_{j=1}^N(\omega_i-\langle\omega\rangle)^2>0$, to avoid a trivial solution.

For all numerical experiments we will consider two network models: Erd\H{o}s-R\'{e}nyi (ER) random networks~\cite{Erdos1960} that are constructed such that a directed link $j\to i$ exists with probability $p$, and networks with scale-free (SF) degree distributions $P(k^{\text{in}})\propto(k^{\text{in}})^{-\gamma}$, $P(k^{\text{out}})\propto(k^{\text{out}})^{-\gamma}$ with enforced minimum degree $k^{\text{in}},k^{\text{out}}\ge k_0$ constructed using the configuration model~\cite{Bekessy1972}. The mean degree of ER and SF networks is given by $\langle k^{\text{in}}\rangle=\langle k^{\text{out}}\rangle=p(N-1)$ and $(\gamma-1)k_0/(\gamma-2)$, respectively. Furthermore, we will consider as an illustrative example Kuramoto-type coupling, i.e., $H(\theta)=\sin(\theta)$. Note that in this case, the coupling is not frustrated [i.e., $H(0)=0$] so that the effective natural frequencies of the linearized system are precisely the natural frequencies, $\bm{\tilde{\omega}}=\bm{\omega}$ [see Eq.~(\ref{eq:Theory02})]. One benefit of this choice is that, because $\bm{\tilde\omega}$ is independent of the coupling strength $K$, so is the optimization of the SAF. However, we stress that our results can be applied of other choices of coupling function, for instance Sakaguchi-Kuramoto coupling~\cite{Sakaguchi1986PTP} as well as coupling with higher harmonics~\cite{Bick2011PRL,Skardal2011PRE,Komarov2013PRL}.

\subsection{Oscillator allocation}\label{sec3sub1}
We consider first the case of oscillator allocation. We assume that an underlying network structure is given, and we can freely choose the natural frequency $\omega_i$ at each node. 
In this case, the optimal choice of $\bm{\omega}$ can be found analytically. By considering the expansion $\bm{\omega}=\sum_{j=1}^N\alpha_j\bm{u}^j$ and inserting it into Eq.~(\ref{eq:Theory13}), we find that $J(\bm{\omega},L)$ can be minimized by placing as much weight as possible into the $N$-th singular vector coefficient $\alpha_N$. However, since the collection of left singular vectors $\bm{u}^j$ tend to have non-zero mean, an additional shift is required to ensure that $\langle\omega\rangle=0$. It is straightforward to show then that the optimal choice of natural frequencies is precisely
\begin{align}
\bm{\omega}=\pm\sigma\sqrt{N}\left(\bm{u}^N-\frac{\langle u^N\rangle}{\langle u^1\rangle}\bm{u}^1\right)\Bigg/\sqrt{1-\frac{\langle u^N\rangle^2}{\langle u^1\rangle^2}},\label{eq:Theory14}
\end{align}
where $\langle u^l\rangle=\frac{1}{N}\sum_{j=1}^Nu_j^l$ represents the mean entry of the $l^{\text{th}}$ left singular vector. Thus, the optimal choice of frequencies is proportional to the dominant left singular vector $\bm{u}^N$ with a shift to retain a mean of zero.

We now demonstrate the effectiveness of the optimal solution given by Eq.~(\ref{eq:Theory14}) in comparison to a random allocation of natural frequencies. Specifically, we consider both ER and SF networks of size $N=100$ with mean degree $\langle k\rangle=4$ (for SF networks we let $\gamma=3$) with (i)~the optimal choice given in Eq.~(\ref{eq:Theory14}) with $\sigma=1$ and (ii)~a set of natural frequencies drawn randomly from the standard normal distribution. We consider 50 networks each of type ER and SF, and for each network we simulate Eq.~(\ref{eq:Kuramoto}) over a range of coupling strengths for both the optimal and random set of natural frequencies. In Fig.~\ref{fig1}(a), we plot the resulting synchronization profiles $r$ vs $K$ (averaged over the 50 network realizations), indicating optimal and random natural frequency results in blue circles and red triangles, respectively. We denote results using ER and SF networks with filled and unfilled markers, respectively. The optimal choice of natural frequencies clearly outperforms the random allocation over the entire range of coupling strength, leading to strong synchronization (i.e., $r\approx1$) even for very small $K$.

\subsection{Oscillator arrangement}\label{sec3sub2}
Next we consider oscillator arrangement. Here we assume that an underlying network structure is given, but rather than choosing natural frequencies freely (i.e., as in Sec.~\ref{sec3sub1}), we are given a set of pre-chosen natural frequencies $\{\omega_i\}$ that must be arranged on the network. In general, the optimal solution depends on the particular network structure and set of natural frequencies given, making analytical treatment impossible. Specifically, since there are $N!$ possible arrangements, it is infeasible to conduct an exhaustive search for the best arrangement even for a moderately sized network. We thus develop a computationally feasible solution for minimizing the SAF.

We propose here a simple greedy algorithm to produce an approximation of the optimal solution. Initially, we arrange the set of natural frequencies (here chosen from the standard normal distribution) randomly on the network. Then, in each step, we propose a switch of a single pair of oscillators, $\omega_i\leftrightarrow\omega_j$, obtain a new frequency vector $\bm{\omega}'$, and compute the new SAF $J(\bm{\omega}',L)$. If the new SAF is less than the previous SAF, $J(\bm{\omega}',L)<J(\bm{\omega},L)$, then the switch is accepted, and otherwise it is rejected. This process is repeated until a chosen number $S$ of switches are proposed. We note that the initial arrangement of oscillators need not be random, and in fact we can start closer to an optimal arrangement by setting an initial arrangement that aligns closely with one or more of the dominant left singular vectors of $L$, saving a significant number of proposed switches.

We demonstrate the power of this approach by simulating the dynamics of Eq.~(\ref{eq:Kuramoto}) on ER and SF networks before and after the natural frequencies are rearranged using the algorithm above. We consider again networks of size $N=1000$ with mean degree $\langle k\rangle=4$ and $\gamma=3$, first arranging normally distributed natural frequencies randomly, then rearranging them with $S=10^5$ proposed switches. We plot the resulting synchronization profiles in Fig.~\ref{fig1}(b), indicating the random and rearranged natural frequency results with blue circles and red triangles, respectively. We again denote ER and SF networks with filled and unfilled markers. As in the case of oscillator allocation, the system with rearranged natural frequencies significantly outperforms the system with randomly arranged natural frequencies. Comparing to the results of optimal allocation that are shown in Fig.~\ref{fig1}(a), in the case of pre-chosen natural frequencies the abrupt transition to synchronization occurs at a larger coupling strength (i.e., $K\approx 0.3$) and the order parameters are generally smaller. Nevertheless, as in the case of optimal arrangement, the order parameter curves in both Fig.~\ref{fig1}(b) significantly outperforms their random counterparts.

\subsection{Network construction}\label{sec3sub3}
We now consider the problem of network construction. We assume that a set of pre-chosen natural frequencies is given, and a network must be build to best synchronize the oscillators. We will assume that a fixed number $M$ of directed links can be made. As in the case of oscillator arrangement, the problem depends sensitively on the particular set of natural frequencies given. Furthermore, there is a combinatorially large number of possible networks without copied or self links that can be built, ${N(N-1) \choose M}$, which makes a full search unfeasible. Thus, we again proceed algorithmically.

We implement an accept-reject algorithm that is initialized with a random network. We begin by constructing a random network with $M$ links around the pre-chosen natural frequencies, which we ensure to be strongly-connected. Next, we propose link rewiring: uniformly at random, we select a link $j\to i$ to delete, and we replace it with a new link between two previously disconnected nodes $j'\to i'$, which are also selected uniformly at random. This yields a new directed Laplacian matrix, $L'$, and new SAF, $J(\bm{\omega},L')$. If the new SAF is less than the previous SAF, $J(\bm{\omega},L')<J(\bm{\omega},L)$, then the link replacement is accepted, and otherwise it is rejected. As before, this process is repeated until $S$ proposed rewirings are considered.

We highlight the effectiveness of this method by simulating the dynamics of Eq.~(\ref{eq:Kuramoto}) on networks before and after the iterative rewiring process. We initialize the rewiring algorithm with both ER and SF networks of size $N=500$ and mean degree $\langle k\rangle=4$ and $\gamma=3$ with normally distributed frequencies, and propose $S=10^4$ link rewirings. We plot the resulting synchronization profiles in Fig.~\ref{fig1}(c), indicating the random and rewired network results with blue circles and red triangles, respectively. We denote the results using initialized ER and SF networks with filled and unfilled markers, respectively. As expected, the networks that are rewired to minimize the SAF outperform the initial networks. Similar to the oscillator arrangement case, the transition to synchronization occurs at a larger coupling strength [i.e., $K\approx0.3$] than in the oscillator allocation case, and it remains abrupt. We note that since the rewiring algorithm is greedy, in principle, the initial network structure may have an effect on the final network. However, we find that initializing the algorithm with ER and SF networks yields very similar results, which suggests that the network initialization does not have a significant effect on the outcome of the rewiring process.

\section{Effect of structural properties of directed networks}\label{sec4}
Having demonstrated the effectiveness of the generalized SAF for optimizing a network's synchronization properties, we next investigate the role of directed network structure in the optimization of synchronization. In particular, we consider two general questions. First, what are the effects {\it of} various directed network structures {\it on} optimal synchronization? Second, what are the effects {\it of} optimal synchronization {\it on} directed network structures? We address these questions below in Secs.~\ref{sec4sub1} and \ref{sec4sub2}, respectively.

\subsection{Degree assortativity}\label{sec4sub1}
We begin by investigating the effects that directed network properties have on optimal synchronization. Perhaps the property that differentiates directed from undirected networks is the characterization of each node by two degrees rather than one (specifically, the in- and out-degrees $k_i^{\text{in}}$ and $k_i^{\text{out}}$ for a given node $i$). Thus, a given directed network can easily be classified with a particular degree assortativity, measuring the correlation between in- and out-degrees at each node in the network. This can be measured with the degree assortativity coefficient~\cite{Newman2003PRE}, i.e., the Pearson correlation coefficient, defined as
\begin{align}
c_{k^{\text{in}},k^{\text{out}}}=\frac{\sum_{i}(k_i^{\text{in}}-\langle k\rangle)(k_i^{\text{out}}-\langle k\rangle)}{\sqrt{\left[\sum_{i}(k_i^{\text{in}}-\langle k\rangle)^2\right]\left[\sum_{i}(k_i^{\text{out}}-\langle k\rangle)^2\right]}},\label{eq:corr0}
\end{align}
where $-1\le c_{k^{\text{in}},k^{\text{out}}}\le1$ such that $c_{k^{\text{in}},k^{\text{out}}}>0$ and $c_{k^{\text{in}},k^{\text{out}}}<0$ indicate a positive and negative correlation, respectively, and $c_{k^{\text{in}},k^{\text{out}}}\approx0$ indicates no correlation. The results presented in Sec.~\ref{sec3} and Fig.~\ref{fig1} used random networks with no correlations between in- and out-degrees. This begs the question: what is the effect of degree assortativity on optimal synchronization?

\begin{figure}[t]
\centering
\epsfig{file =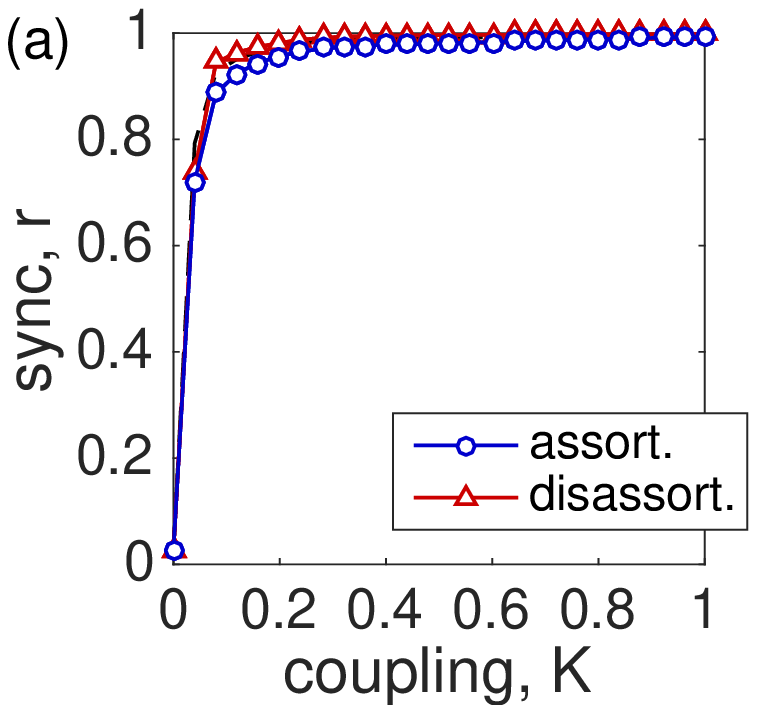, clip =,width=0.49\linewidth }
\epsfig{file =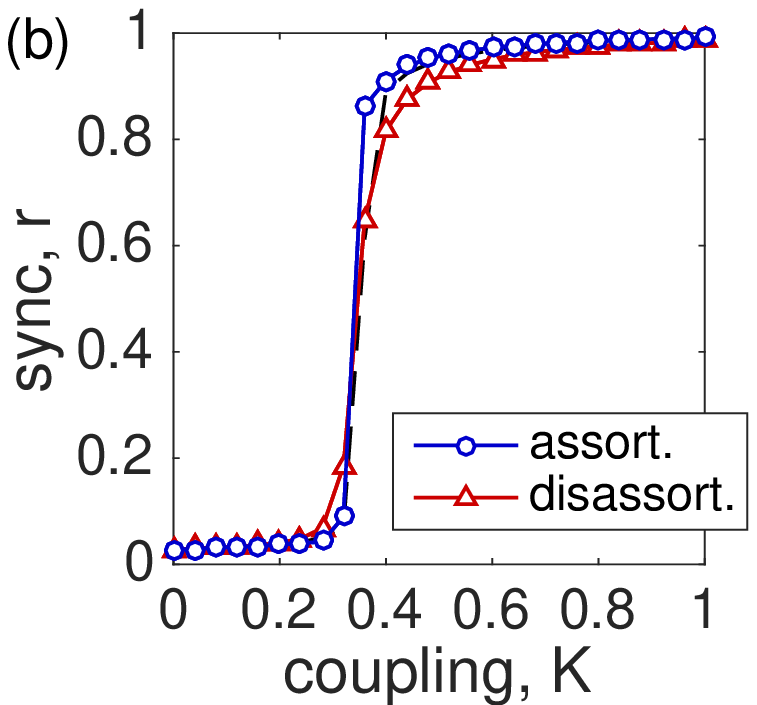, clip =,width=0.49\linewidth }
\caption{(Color online) Effect of degree assortativity (ER networks): Synchronization profiles $r$ vs $K$ for optimally (a) allocated and (b) arranged frequencies on ER networks with $N=1000$, $\langle k\rangle=4$ (dashed curves) as well as these networks rewired to be assortative with $c_{k^{\text{in}},k^{\text{out}}}=0.99$ (blue circles) and disassortative with $c_{k^{\text{in}},k^{\text{out}}}=-0.92$ (red triangles). Results represent an average over 50 network realizations, and the frequencies are normally distributed for (b). Uncorrelated results are plotted with a dashed curve.}
\label{fig2}
\end{figure}

\begin{figure}[t]
\centering
\epsfig{file =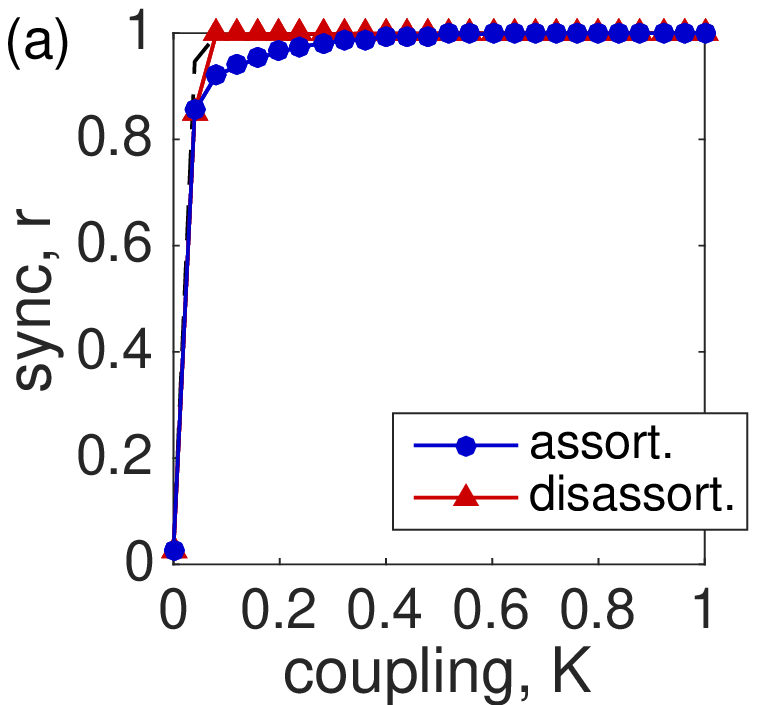, clip =,width=0.49\linewidth }
\epsfig{file =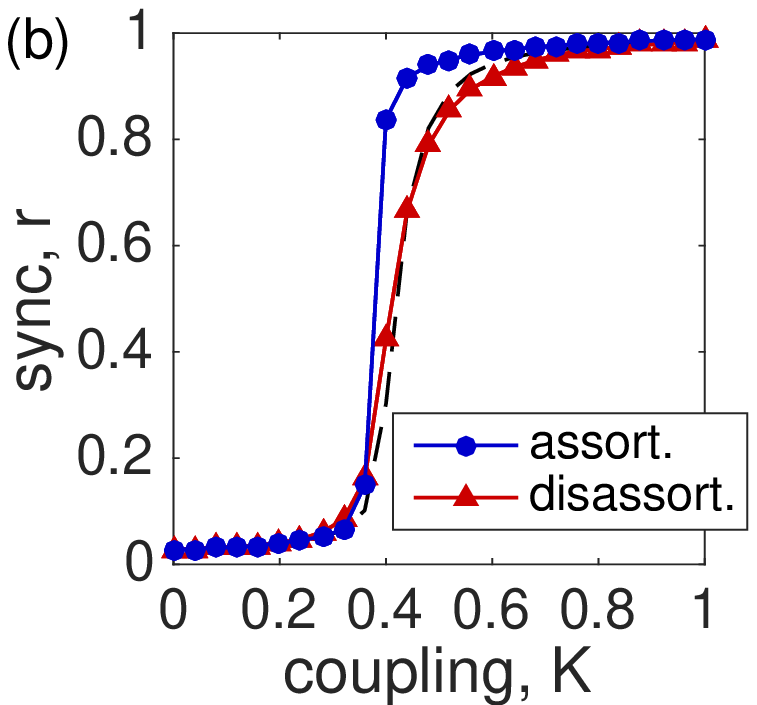, clip =,width=0.49\linewidth }
\caption{(Color online) Effect of degree assortativity (SF networks): Synchronization profiles $r$ vs $K$ for optimally (a) allocated and (b) arranged frequencies on SF networks with $N=1000$, $\langle k\rangle=4$, $\gamma=3$ (dashed curves) as well as these networks rewired to be assortative with $c_{k^{\text{in}},k^{\text{out}}}=0.94$ (blue circles) and disassortative with $c_{k^{\text{in}},k^{\text{out}}}=-0.19$ (red triangles). Results represent an average over 50 network realizations, and the frequencies are normally distributed for (b). Uncorrelated results are plotted with a dashed curve.}
\label{fig3}
\end{figure}

We investigate the effects of assortative and disassortative degree correlations on optimal synchronization in the cases of oscillator allocation and arrangement. In particular, given the sequence of in- and out-degrees for a particular network, we construct two new networks using in- and out-degree pairs set to maximize and minimize the degree assortativity. Maximum assortativity is obtained by matching the maximum in-degree with the maximum out degree, etc., and maximum disassortativity is obtained by matching the maximum in-degree with the minimum out-degree, etc. Beginning with ER networks of size $N=1000$ with mean degree $\langle k\rangle=4$, we plot the synchronization profiles $r$ vs $K$ of the original network (dashed curves) and its assortative (blue circles) and disassortative (red triangles) rewirings after optimally allocating and arranging frequencies in Figs.~\ref{fig2}(a) and (b), respectively. The results represent an average over $50$ network realizations, and in the case of oscillator arrangement, the frequencies are normally distributed. We repeat these simulations on SF networks of size $N=1000$ with $\langle k\rangle=4$ and $\gamma=3$, plotting the results from optimal allocation and arrangement in Figs.~\ref{fig3}(a) and (b), respectively.

We note first that, while there is a difference between optimal synchronization in assortative and disassortative networks, the optimization methods presented above are effective in both cases. Interestingly, we find that the effect of degree assortativity on synchrony optimization differs between the cases of optimal allocation and optimal arrangement. In particular, we observe that in both ER and SF networks when oscillators are optimally allocated, disassortative networks tend to outperform assortative networks [see Fig.~\ref{fig2}(a) and Fig.~\ref{fig3}(a)]. However, when oscillators are optimally arranged, assortative networks tend to outperform disassortative networks [see Fig.~\ref{fig2}(b) and Fig.~\ref{fig3}(b)]. Thus, we conclude that positive/negative degree assortativity in directed networks does not necessarily imply improved or diminished synchronization properties, and depends on the nature of the constraints on the optimization problem. However, we do point out that these tendencies do not appear to depend much on network heterogeneity, since the results for ER and SF networks are similar.

\begin{figure*}[t]
\centering
\epsfig{file =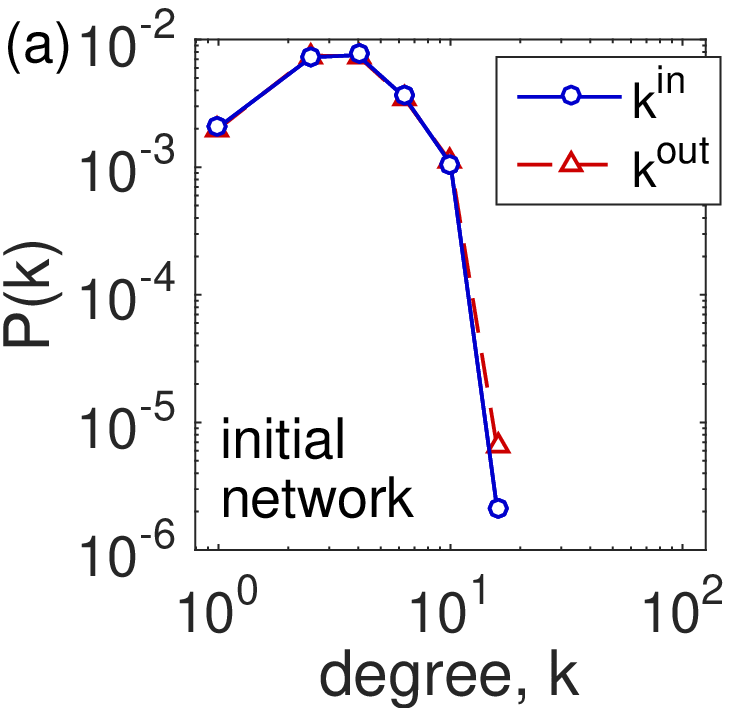, clip =,width=0.25\linewidth }
\epsfig{file =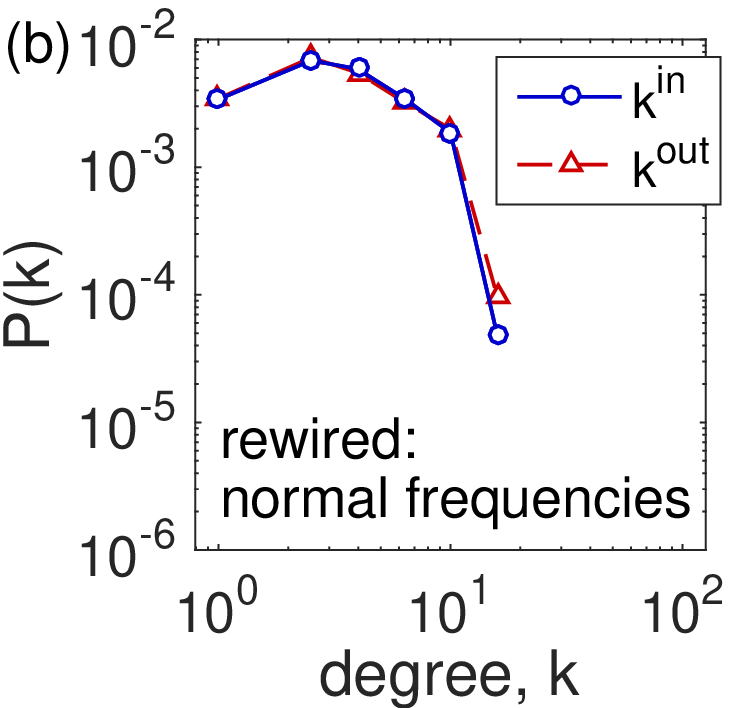, clip =,width=0.25\linewidth }
\epsfig{file =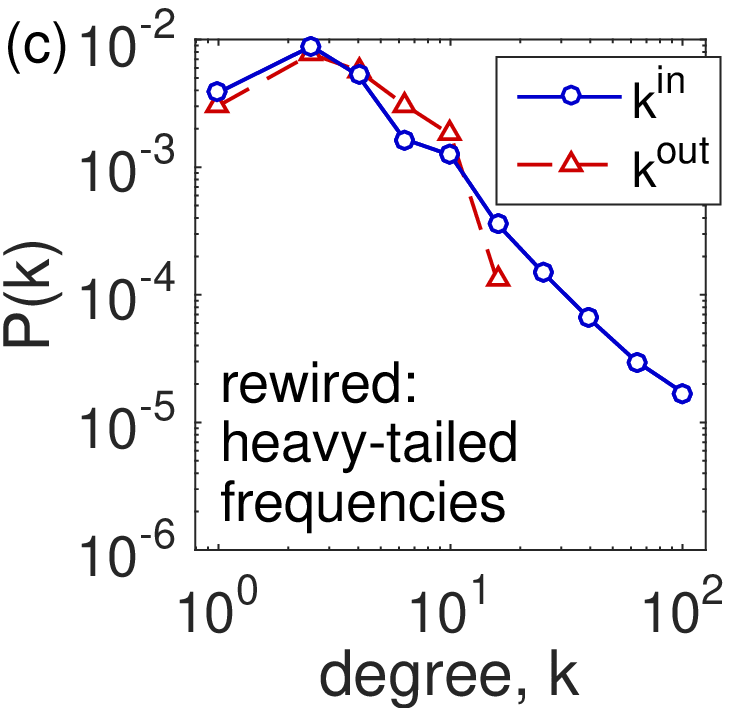, clip =,width=0.25\linewidth }
\caption{(Color online) Effect of optimization on directed network structure: Degree distributions $P(k)$ of networks (a) before and (b), (c) after rewiring for optimal synchronization. In (b) the natural frequencies are normally distributed, while in (c) the frequency distribution is heavy-tailed as defined in Eq.~(\ref{eq:heavy}). The initial network is ER with $N=500$ and $\langle k\rangle=4$. Results represent an average over 50 network realizations.}\label{fig4}
\end{figure*}

\subsection{Evolution of network structure}\label{sec4sub2}
Next, we investigate the effect that optimization has on a given directed network's structure. In particular, we consider the case of network rewiring and study the evolution of the degree distribution $P(k)$ for both in- and out degrees $k^{\text{in}}$ and $k^{\text{out}}$. We proceed as follows. We initialize an ER network of size $N=500$ with $\langle k\rangle=4$ and a set of pre-chosen natural frequencies. After noting the initial degree distributions of $k^{\text{in}}$ and $k^{\text{out}}$, we apply the rewiring algorithm described in Sec.~\ref{sec3sub3} for $S=5\cdot10^4$ proposed rewirings and study the resulting degree distributions of $k^{\text{in}}$ and $k^{\text{out}}$. Importantly, we consider two classes of natural frequencies. First, we consider frequencies drawn from the standard normal distribution (as used above), representing a homogeneous collection of frequencies. Second, we consider frequencies drawn from a much broader, heavy-tailed distribution described by
\begin{align}
P(\omega)=\left\{\begin{array}{cl}(\beta-1)/2\beta\omega_0 & \text{if }|\omega|<\omega_0,\\ ((\beta-1)\omega_0^{\beta-1}/2\beta\omega_0) |\omega|^{-\beta} & \text{if }|\omega|\ge\omega_0,\end{array}\right.\label{eq:heavy}
\end{align}
which represents a symmetric (continuous and piecewise-smooth) distribution on $\mathbb{R}$ that is uniform for $|\omega|\le\omega_0$ and has a power-law decay with exponent $\beta$ for $|\omega|>\omega_0$. We choose $\beta=2.5$ and $\omega_0=1$.

We present results for this experiment in Fig.~\ref{fig4}, plotting the degree distributions $P(k^{\text{in}})$ and $P(k^{\text{out}})$ for the initial networks in panel (a) in blue circles and red triangles, respectively, and the resulting distributions after rewiring for optimal synchronization for the cases of normal and heavy-tailed frequencies in panels (b) and (c), respectively. Results represent an average over $50$ network realizations. We emphasize that the initial networks used for the cases of normal vs heavy-tailed frequencies are identical, and thus the difference in the distributions after rewiring for optimality are due solely to the effect that the different frequency distributions have on the rewiring process. Noting that results are plotted in log-log format, both the in- and out-degree distributions of the initial networks are thin and in fact, they remain thin after rewiring for the case of normal frequencies. However, in the case of heavy-tailed frequencies we observe that while the out-degree distribution remains thin, the in-degree distribution becomes very wide, indicating the emergence of nodes with very large in-degree compared to the rest of the network, or hubs. 

These results shed some light on the role of directedness of networks in optimal synchronization. In the case of undirected networks, we found\cite{Skardal2014PRL} in a similar rewiring experiment that the heterogeneity of the degree distribution of an optimized network roughly matches the heterogeneity of the frequency distribution -- heavy-tailed frequency distributions give rise to heavy-tailed degree distributions, and thin frequency distributions give rise to thin degree distributions. The results for directed networks provide an interesting contrast to this phenomenon. In particular, the heterogeneity of the in-degree distribution matches the heterogeneity of the frequency distribution, but no such relationship is observed for the out-degree distribution. This suggests that in the case of directed networks, in-degrees play an important role in optimizing synchronization properties, while out-degrees are much less significant. As we will see below, the role of in- and out-degrees extend to the correlations we observe between the degrees and frequencies. 

\section{Effect of Interplay Between Structural and Dynamical Properties}\label{sec5}
Having investigated the effects of directed network structure on optimal synchronization, and vice versa, we now consider the relationship between the structural and dynamical properties of optimal networks. In particular, \emph{what} relationships between dynamical and structural properties promote synchronization in directed networks? In our previous work~\cite{Skardal2014PRL}, we observed two general properties that are common to undirected synchrony-optimized networks: (i) a strong positive correlation between an oscillator's nodal degree and its natural frequency, and (ii) a strong negative correlation between an oscillator's natural frequency and the average natural frequencies of its network neighbors. These results are consistent with those found to promote global synchronization in other studies~\cite{Brede2008PLA,Buzna2009PRE,Kelly2011Chaos,Scafuti2015PRE,Pinto2015PRE}; however, in each of these cases only undirected networks were considered. This leaves a significant gap in our understanding of which directed network properties promote synchronization. 

In Sec.~\ref{sec4sub2}, we observed that in-degrees play a particularly important role in synchrony optimization compared to out-degrees. This begs the question: How does the relationship between in-degrees and frequencies compare to the relationship between out-degrees and frequencies? To address this question, we study the system properties of synchrony-optimized networks resulting from the optimization of the general SAF. In Sec.~\ref{sec5sub1}, we study correlations between oscillators' natural frequencies and their nodal in- and out-degrees. In Sec.~\ref{sec5sub2}, we study correlations between the natural frequencies of neighboring oscillators.

\subsection{Degree-frequency correlations}\label{sec5sub1}
We begin our investigation by considering correlations between each oscillator's nodal in- and out-degree, $k_i^{\text{in}}$ and $k_{i}^{\text{out}}$, with their respective natural frequency in absolute value, $|\omega_i|$. We present results for synchrony-optimized networks arising for the optimization problem of rearranging a set of pre-chosen frequencies on a given network (see Sec.~\ref{sec3sub2}); however, we emphasize that these results are in close agreement with our study of networks resulting for the other optimization problems discussed in Sec.~\ref{sec3}.

\begin{figure}[t]
\centering
\epsfig{file =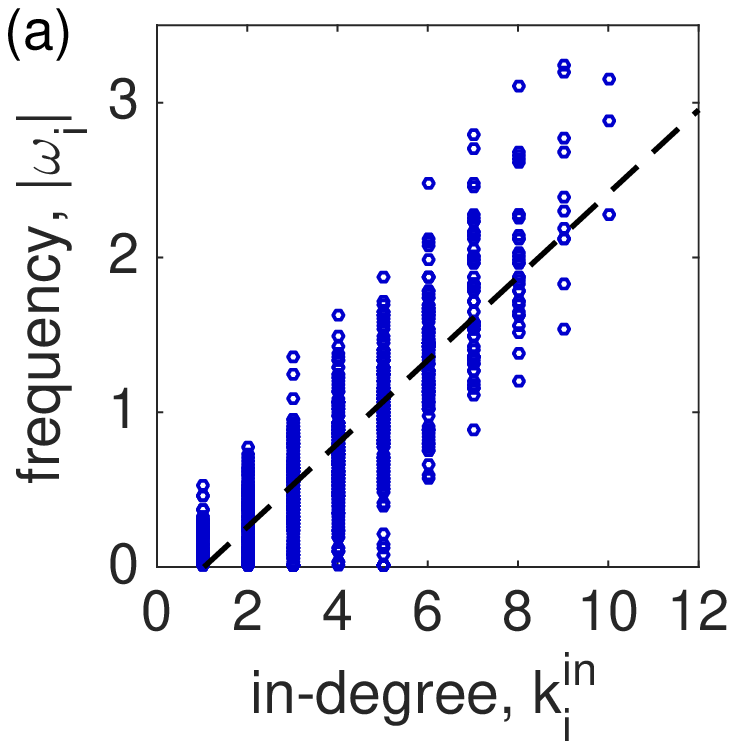, clip =,width=0.49\linewidth }
\epsfig{file =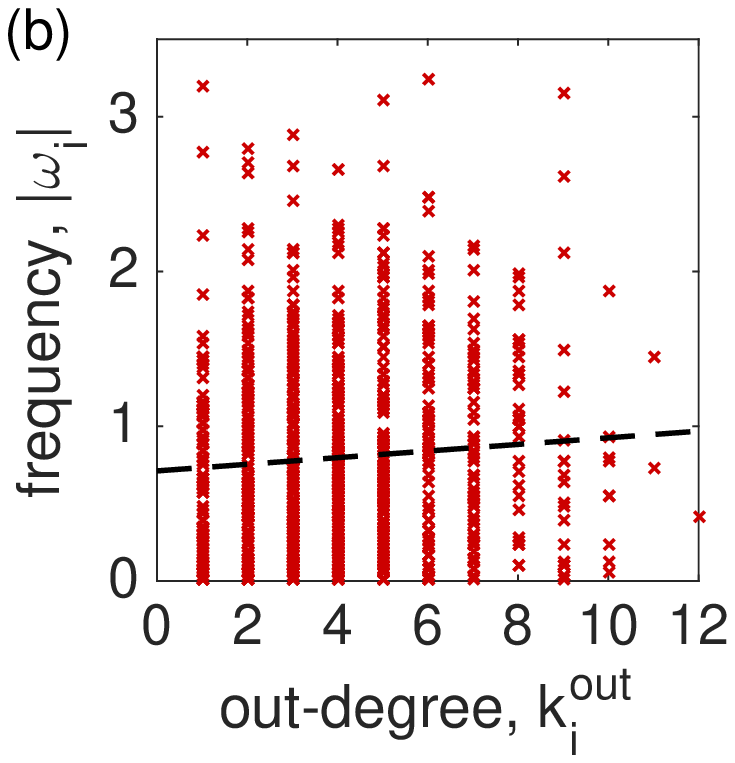, clip =,width=0.49\linewidth }
\caption{(Color online) Degree-frequency correlations: For an ER network of size $N=1000$ with $\langle k\rangle=4$ and normally-distributed natural frequencies, the natural frequency $|\omega_i|$ in magnitude vs (a) the nodal in-degree $k_i^{\text{in}}$ and (b) nodal out-degree $k_i^{\text{out}}$. The least squares line of best fit is plotted in dashed black, and the Pearson correlation coefficients for the results in panels (a) and (b) are $c_{k^{\text{in}},|\omega|}=0.8530$ and $c_{k^{\text{out}},|\omega|}=0.0726$, respectively.}\label{fig5}
\end{figure}

In Fig.~\ref{fig5}, we show a scatter plots of $|\omega_i|$ versus (a)~$k_i^{\text{in}}$ and (b) $k_i^{\text{out}}$. Black dashed lines indicate the least squares line of best fit for each case. We show results for an example ER network of size $N=1000$ with mean degree $\langle k\rangle=4$ and natural frequencies that are drawn from the standard normal distribution. The natural frequencies are arranged on the network following the algorithm described in Sec.~\ref{sec3sub2} with $S=10^5$ proposed switches. By comparing Fig.~\ref{fig5}(a) and (b), we observe contrasting relationships that natural frequencies have with in-degrees and out-degrees. In particular, a strong positive relationship is clear between natural frequencies and in-degrees, but no such relationship is clear between natural frequencies and out-degrees. To support this observation, we calculate the assortativity coefficient between the respective quantities,
\begin{align}
c_{k,|\omega|}=\frac{\sum_{i}(k_i-\langle k\rangle)(|\omega_i|-\langle|\omega|\rangle)}{\sqrt{\left[\sum_{i}(k_i-\langle k\rangle)^2\right]\left[\sum_{i}(|\omega_i|-\langle|\omega|\rangle)^2\right]}}.\label{eq:corr2}
\end{align}
We find that in our particular example, $c_{k^{\text{in}},|\omega|}=0.8530$ and $c_{k^{\text{out}},|\omega|}=0.0726$.

\subsection{Frequency-frequency correlations}\label{sec5sub2}
Next, we consider the relationship between natural frequencies of neighboring oscillators. We note that in the directed case, this relationship is more nuanced than in the undirected case, and therefore we consider the relationships between a given oscillator's natural frequency and the mean natural frequency of its neighbors (i) along in-coming links and (ii) along out-going links. To investigate these relationships, we define the mean neighboring frequencies, respectively, as
\begin{align}
\langle\omega\rangle_i^{\text{in}}=\frac{1}{k_i^{\text{in}}}\sum_{j=1}^NA_{ij}\omega_j,\hskip2ex\langle\omega\rangle_i^{\text{out}}=\frac{1}{k_i^{\text{out}}}\sum_{j=1}^NA_{ji}\omega_j.\label{eq:corr1}
\end{align}
Physically, the mean in-frequency for oscillator $i$, $\langle\omega\rangle_i^{\text{in}}$ represents the mean frequency of all the oscillators that {\it influence} oscillator $i$, while the mean out-frequency for oscillator $i$, $\langle\omega\rangle_i^{\text{in}}$ represents the mean frequency of all the oscillators that are {\it influenced by} oscillator $i$. 

\begin{figure}[t]
\centering
\epsfig{file =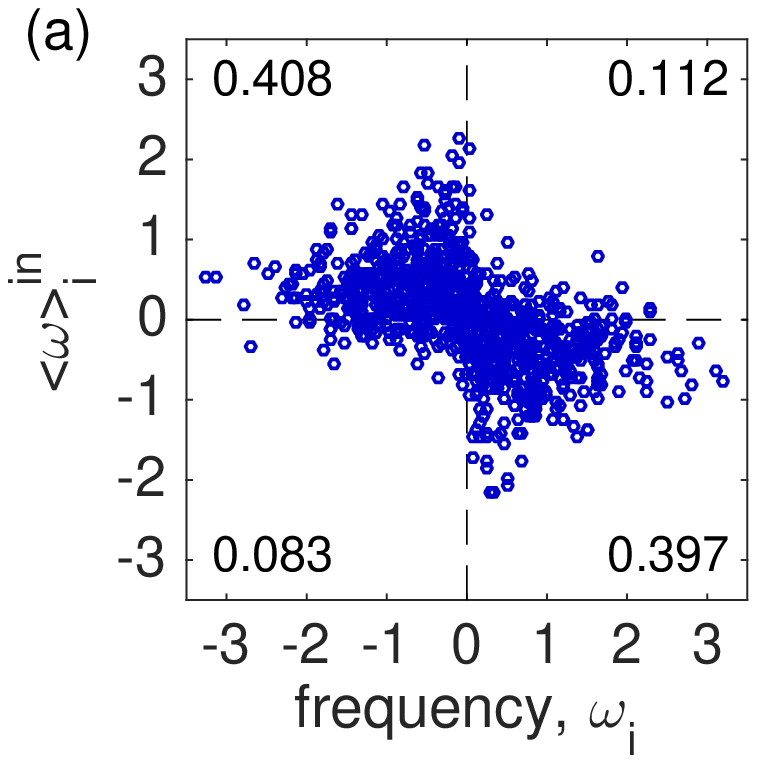, clip =,width=0.49\linewidth }
\epsfig{file =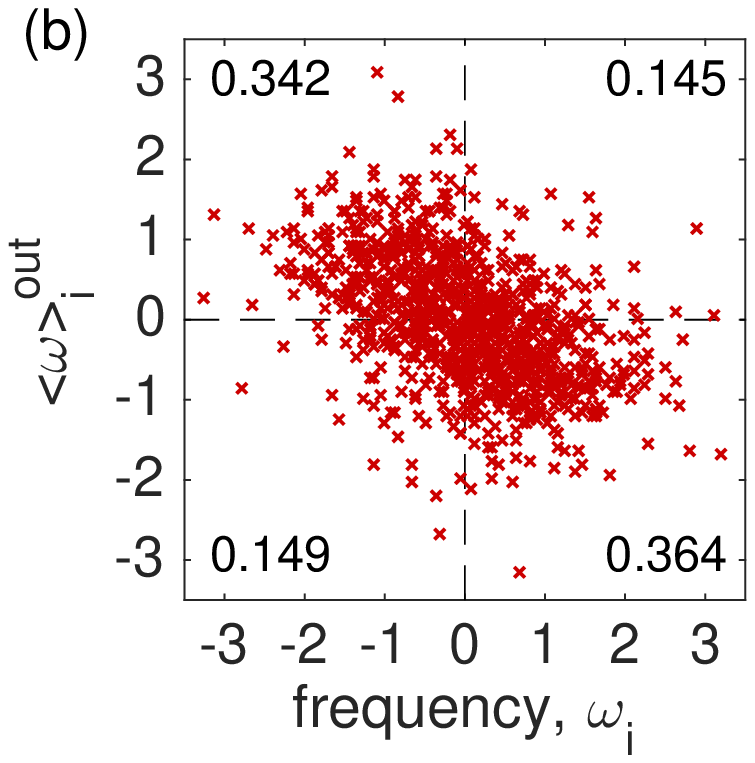, clip =,width=0.49\linewidth }
\caption{(Color online) Frequency-frequency correlations: For an ER network of size $N=1000$ with $\langle k\rangle=4$ and normally-distributed natural frequencies, the mean (a) in-frequency $\langle\omega\rangle_i^{\text{in}}$ and (b) out-frequency $\langle\omega\rangle_i^{\text{out}}$ vs the oscillator's natural frequency $\omega_i$. Note the ``bow-tie'' structure in panel (a). Numbers in each corner indicate the fraction of $(\omega_i,\langle\omega\rangle_i)$ pairs that fall in each respective quadrant.}\label{fig6}
\end{figure}

Using the same network as in the previous example, we plot $\langle\omega\rangle_i^{\text{in}}$ and $\langle\omega\rangle_i^{\text{out}}$ in Fig.~\ref{fig6}(a) and (b), respectively, against the natural frequency $\omega_i$. In both panels, one can observe a negative relationship. However, the two cases differ in their structure -- note the strong ``bow-tie''-like structure in Fig.~\ref{fig6}(a). This bow-tie structure is reminiscent of the frequency-frequency relationship that we previously observed for undirected networks~\cite{Skardal2014PRL}. To quantify this structure, we count the fraction of pairs $(\omega_i,\langle\omega\rangle_i)$ that fall into each quadrant, which we indicate in the respective corners of Fig.~\ref{fig6}(a) and (b). This allows us to quantify the proportion of oscillators whose natural frequency shares their sign with the frequencies opposite incoming and outgoing links. Approximately $20\%$ of the natural frequencies share a sign with their respective mean in-frequency, while $30\%$ share a sign with their respective mean out-frequency. This finding suggests that the relationship between each natural frequency and its respective mean in-frequency has a more significant effect on synchronization than the relationship with its respective mean out-frequency.

\subsection{Evolution of correlations}\label{sec5sub3}
Finally, we investigate the evolution of a network's properties during the optimization process. Again, we restrict our attention to oscillator arrangement, as discussed in Sec.~\ref{sec3sub2}, and we thus consider the evolution of a network through a process of switching the natural frequencies. Throughout this evolution, we will consider the correlation between both in- and out-degrees and natural frequencies, measured by the assortativity coefficient in Eq.~(\ref{eq:corr2}), as well as frequency-frequency correlations measured by
\begin{align}
c_{\omega,\omega}=\frac{\sum_{i,j}A_{ij}(\omega_i-\langle\omega\rangle)(\omega_j-\langle\omega\rangle)}{\sqrt{\left(\sum_{i,j}A_{ij}(\omega_i-\langle\omega\rangle)^2\right)\left(\sum_{i,j}A_{ij}(\omega_j-\langle\omega\rangle)^2\right)}},\label{eq:corr3}
\end{align}
which in this case quantifies the assortativity between neighboring natural frequencies.

\begin{figure}[t]
\centering
\epsfig{file =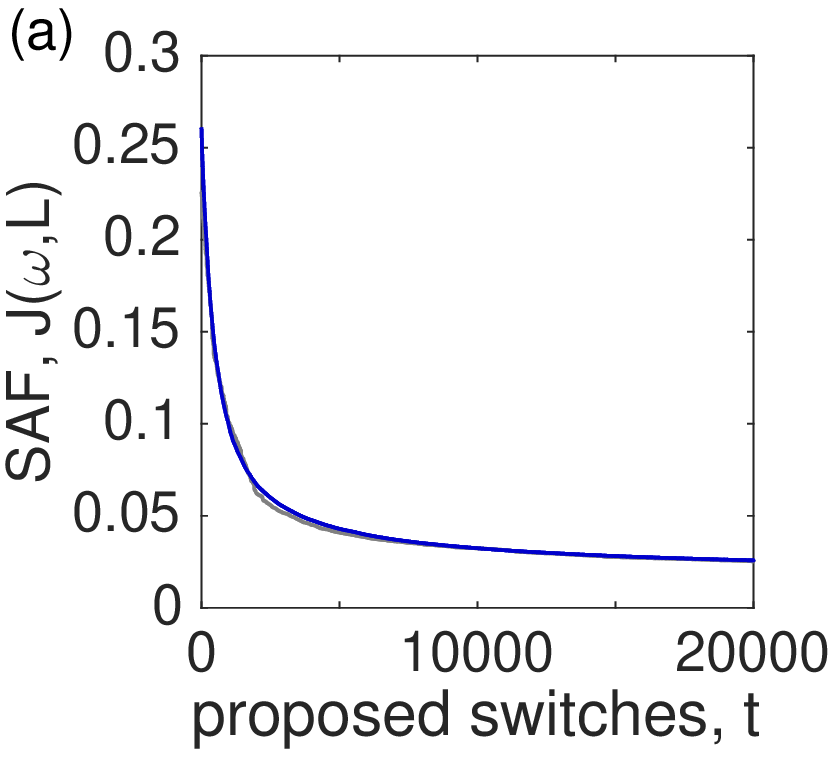, clip =,width=0.49\linewidth }
\epsfig{file =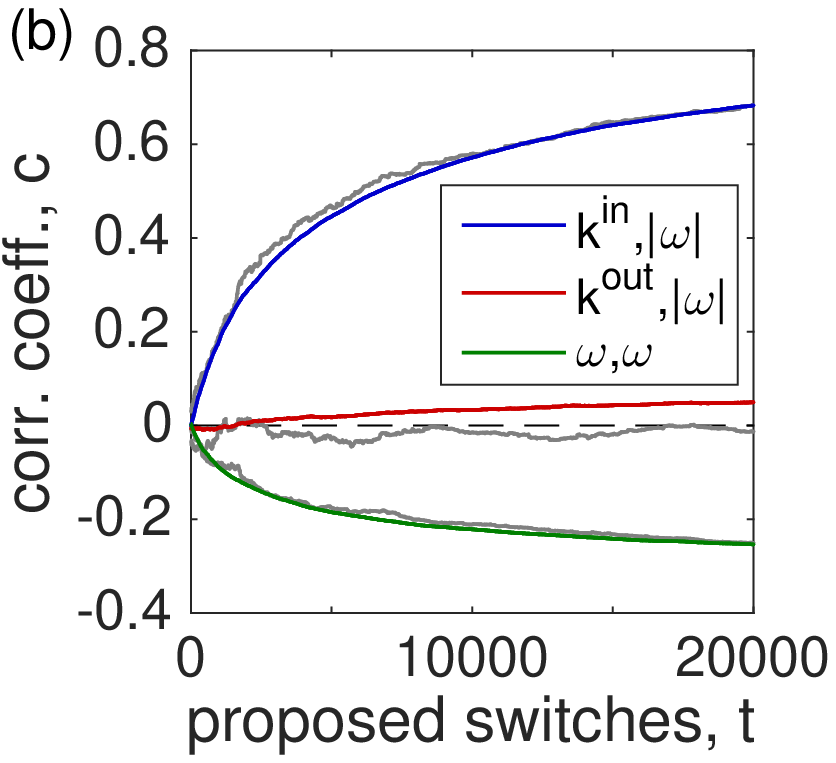, clip =,width=0.49\linewidth }
\caption{(Color online) Evolution of correlations: Averaged over $50$ ER networks of size $N=1000$ with $\langle k\rangle=4$ and normally-distributed natural frequencies, the evolution of (a) the SAF $J(\bm{\omega},L)$ and (b) the correlation coefficients $c_{k^{\text{in}},|\omega|}$ (top, blue), $c_{k^{\text{out}},|\omega|}$ (middle, red), and $c_{\omega,\omega}$ (bottom, green) as natural frequency switches are proposed. Gray curves indicate results for a single realization.}\label{fig7}
\end{figure}

In Fig.~\ref{fig7}, we plot the results obtained from optimally rearranging natural frequencies drawn from a standard normal distribution on an ER network of size $N=1000$ with mean degree $\langle k\rangle=4$ over the course of $S=2\times10^4$ proposed switches. For context, we plot the evolution of the SAF $J(\bm{\omega},L)$ in Fig.~\ref{fig7}(a). In Fig.~\ref{fig7}(b), we plot the correlation coefficients $c_{k^{\text{in}},|\omega|}$, $c_{k^{\text{out}},|\omega|}$, and $c_{\omega,\omega}$ in blue, red and green, respectively. (We note that these also align from top to bottom, respectively.) Results represent an average over $50$ network realizations. For comparison, results for a single realization are plotted as gray curves. 

Our main observation is that as more switches are proposed, we see a rapid shift in the SAF, $c_{k^{\text{in}},|\omega|}$, and $c_{\omega,\omega}$, but only a slight shift in $c_{k^{\text{out}},|\omega|}$. As the network's synchronization properties improve, the correlation coefficients $c_{k^{\text{in}},|\omega|}$ and $c_{\omega,\omega}$ quickly saturate to approximately $0.7$ and $-0.3$, respectively. In contrast, $c_{k^{\text{out}},|\omega|}$ remains small in comparison, but is on average positive. These findings are consistent with the numerical results presented in Secs.~\ref{sec5sub1} and \ref{sec5sub2} on the degree-frequency correlations and frequency-frequency correlations observed in optimally rewired networks.

\section{Discussion}\label{sec6}
In this paper, we have developed a framework for optimizing the synchronization properties of directed networks of coupled oscillators. Our results can be regarded as an extension to previous work that focused on the case of undirected networks~\cite{Skardal2014PRL}. Our main theoretical result is the derivation of the generalized synchrony alignment function given by Eq.~\eqref{eq:Theory13}, which we have shown can be used to systematically optimize the synchronization properties of networks under several constraints, including oscillator allocation (Sec.~\ref{sec3sub1}), oscillator arrangement (Sec.~\ref{sec3sub2}), and network construction (Sec.~\ref{sec3sub3}). We emphasize that this approach is efficient, does not require large-scale simulation of dynamics [i.e., either Eq.~\eqref{eq:Kuramoto} or Eq.~\eqref{eq:Theory01}], and is based on an objective measure of synchronization (i.e., a perturbation analysis of the Kuramoto order parameter) and not heuristics.

The generalized SAF approach presented here shows that the synchronization properties of a general directed network depends on the alignments of the natural frequency vector with the left singular vectors of the Laplacian, weighted appropriately by the left singular values. In particular, stronger synchronization is attained as the natural frequency vector becomes aligned with dominant singular vectors, i.e., those associated with larger singular values. This is a natural generalization from the undirected case, where it was found that stronger alignments of the natural frequency vector with the more dominant eigenvectors of the Laplacian promote synchronization. We emphasize that our approach is designed to maximize the Kuramoto order parameter in the regime where strong synchronization may occur, but point out that even for relatively small coupling strengths the approach works remarkably well. We also point out that the method is not designed to minimize the critical coupling strength associated to the onset of  synchronization. Furthermore, we hypothesize that our results may be applied to better understand the effect that failures and other perturbations have on synchronization in complex networks~\cite{Witthaut2016PRL}.

To provide further insight into the mechanisms that enhance synchronization and the role of directedness, we investigated the effects that directed network properties have on optimal synchronization, the effects that optimization has on directed network structures. In undirected networks, it has been observed that after rewiring for optimal synchronization the heterogeneity of the degree distribution roughly matches the heterogeneity of the frequency distribution. We have found a contrasting phenomenon for directed networks: after rewiring for optimal synchronization the heterogeneity of the in-degree distribution roughly matches the heterogeneity of the frequency distribution; no such relationship was found for the out-degrees. We have also studied the relationship between structural and dynamical properties of optimized networks. In the undirected case, it is well known that a positive correlation between degrees and natural frequencies and a negative correlation between neighboring oscillators' frequencies promote synchronization. We have shown that these relationships become more nuanced in the directed case; synchronization is promoted by a positive correlation between the nodal in-degree of an oscillator and its respective natural frequency, while the relationship between the nodal out-degree of an oscillator and its respective natural frequency has comparatively little effect. We also studied correlations between the natural frequencies of neighboring oscillators and found that directed synchrony-optimized networks display a stronger relationship between each natural frequency and the mean natural frequency of its neighbors along in-coming links than its neighbors along out-going links. Taken together, these results indicate that in-degrees play a significant role in the synchronization properties of directed networks, while out-degrees are much less significant.

\acknowledgments
DT acknowledges support from NIH Award No. R01HD075712. JS acknowledges support from Simons Foundation Grant No. 318812 and Army Research Office Grant No. W911NF-12-1-0276.

\bibliographystyle{plain}

\begin{thebibliography}{99}
\bibitem{Strogatz2003} S. H. Strogatz, {\it Sync: the Emerging Science of Spontaneous Order} (Hypernion, 2003).
\bibitem{Pikovsky2003} A. Pikovsky, M. Rosenblum, and J. Kurths, {\it Synchronization: A Universal Concept in Nonlinear Sciences} (Cambridge University Press, 2003).
\bibitem{Arenas2008PR} A. Arenas, A. D\'{i}az-Guilera, J. Kurths, Y. Moreno, and C. Zhou, Phys. Rep. {\bf 469}, 93 (2008).
\bibitem{Glass1988} L. Glass and M. C. Mackey, {\it From Clocks to Chaos: The Rhythms of Life} (Princeton University Press, Princeton, 1988).
\bibitem{Yamaguchi2003Science} S. Yamaguchi et al., Science {\bf 302}, 1408 (2003).
\bibitem{Wiesenfeld1996PRL} K. Wiesenfeld, P. Colet, and S. H. Strogatz, Phys. Rev. Lett. {\bf 76}, 404 (1996).
\bibitem{Motter2013NaturePhysics} A. E. Motter, S. A. Myers, M. Anghel, and T. Nishikawa, Nat. Phys. {\bf 9}, 191 (2013).
\bibitem{Brede2008PLA} M. Brede, Phys. Lett. A {\bf 372}, 2618 (2008).
\bibitem{Buzna2009PRE} L. Buzna, S. Lozano, and A. D\'{i}az-Guilera, Phys. Rev. E {\bf 80}, 066120 (2009).
\bibitem{Kelly2011Chaos} D. Kelly and G. A. Gottwald, Chaos {\bf 21}, 025110 (2011).
\bibitem{Scafuti2015PRE} F. Scafuti, T. Aoki, and M. di Bernardo, Phys. Rev. E {\bf 91}, 062913 (2015).
\bibitem{Pinto2015PRE} R. S. Pinto and A. Saa, Phys. Rev. E {\bf 92}, 062801 (2015).
\bibitem{Skardal2014PRL} P. S. Skardal, D. Taylor, and J. Sun, Phys. Rev. Lett. {\bf 113}, 144101 (2014).
\bibitem{KroghMadsen2012} T. Krogh-Madsen and D. J. Christini, Annu. Rev. Biomed. Eng. {\bf 14}, 179 (2012).
\bibitem{Karma2013Rev} A. Karma, Annu. Rev. Condens. Matter Phys. {\bf 4}, 313 (2013).
\bibitem{Prindle2012Nature} A. Prindle, P. Samayoa, I. Razinkov, T. Danino, L. S. Tsimring, and J. Hasty, Nature {\bf 481}, 39 (2012).
\bibitem{Rohden2012PRL} M. Rohden, A. Sorge, M. Timme, and D. Witthaut, Phys. Rev. Lett. {\bf 109}, 064101 (2012).
\bibitem{Dorfler2013PNAS} F. D\"{o}rfler, M. Chertkov, and F. Bullo, Proc. Natl. Acad. Sci. {\bf 110}, 2005 (2013).
\bibitem{Skardal2015SciAdv} P. S. Skardal and A. Arenas, Sci. Adv. {\bf 1}, e1500339 (2015).
\bibitem{Witthaut2016PRL} D. Witthaut, M. Rohden, X. Zhang, S. Hallerberg, and M. Timme, Phys. Rev. Lett. {\bf 116}, 138701 (2016).
\bibitem{Kuramoto1984} Y. Kuramoto, {\it Chemical Oscillations, Waves, and Turbulence} (Springer, New York, 1984).
\bibitem{Restrepo2005Chaos} J. G. Restrepo, E. Ott, and B. R. Hunt, Chaos {\bf 16}, 015107 (2005).
\bibitem{Skardal2015PREb} P. S. Skardal, D. Taylor, J. Sun, and A. Arenas, Phys. Rev. E {\bf 91}, 010802(R) (2015).
\bibitem{Restrepo2005PRE} J. G. Restrepo, E. Ott, and B. R. Hunt, Phys. Rev. E {\bf 71}, 036151 (2005).
\bibitem{GomezGardenes2007PRL} J. G\'{o}mez-Garde\~{n}es, Y. Moreno, and A. Arenas, Phys. Rev. Lett. {\bf 98}, 034101 (2007).
\bibitem{GomezGardenes2011PRL} J. G\'{o}mez-Garde\~{n}es, S. G\'{o}mez, A. Arenas, and Y. Moreno, Phys. Rev. Lett. {\bf 106} 128701 (2011).
\bibitem{Skardal2012PRE} P. S. Skardal and J. G. Restrepo, Phys. Rev. E {\bf 85}, 016208 (2012).
\bibitem{Restrepo2014EPL} J. G. Restrepo and E. Ott, Europhys. Lett. {\bf 107}, 60006 (2014).
\bibitem{Skardal2015PRE} P. S. Skardal, J. G. Restrepo, and E. Ott, Phys. Rev. E {\bf 91}, 060902(R) (2015).
\bibitem{Arenas2006PRL} A. Arenas, A. D\'{i}az-Guilera, and C. J. Per\'{e}z-Vicente, Phys. Rev. Lett. {\bf 96}, 114102 (2006).
\bibitem{Skardal2015} P. S. Skardal, D. Taylor, J. Sun, and A. Arenas, Phys. Rev. E {\bf 93}, 042314 (2016).
\bibitem{Sanchez2002PRL} A. S\'{a}nchez, J. M. L\'{o}pez, and M. A. Rodr\'{i}guez, Phys. Rev. Lett. {\bf 88}, 048701 (2002).
\bibitem{Restrepo2008PRL} J. G. Restrepo, E. Ott, and B. Hunt, Phys. Rev. Lett. {\bf 100}, 058701 (2008).
\bibitem{Lentz2012PRE} H. H. K. Lentz, T. Selhorst, and I. M. Sokolov, Phys. Rev. E {\bf 85}, 066111 (2012).
\bibitem{Pecora1998PRL} L. M. Pecora and T. L. Carroll, Phys. Rev. Lett. {\bf 80}, 2109 (1998).
\bibitem{Nishikawa2006PRE} T. Nishikawa and A. E. Motter, Phys. Rev. E {\bf 73}, 065106 (2006).
\bibitem{Nishikawa2010PNAS} T. Nishikawa and A. E. Motter, Proc. Natl. Acad. Sci. U.S.A. {\bf 107}, 10342 (2010).
\bibitem{Ravoori2011PRL} B.~Ravoori, A.~B.~Cohen, J.~Sun, A.~E.~Motter, T.~E.~Murphy, and R.~Roy, Phys. Rev. Lett. {\bf 107}, 034102 (2011).
\bibitem{BenIsrael1974} A. Ben-Israel and T. N. E. Grenville, {\it Generalized Inverses} (Springer, New York, 1974).
\bibitem{Golub}G. H. Golub and C. F. Van Loan, {\it Matrix Computations} (The John Hopkins University Press, 1996).
\bibitem{Sakaguchi1986PTP}  H. Sakaguchi and Y. Kuramoto, Prog. Theor. Phys. {\bf 76}, 576 (1986).
\bibitem{Bick2011PRL} C. Bick, M. Timme, D. Paulikat, D. Rathlev, and P. Ashwin, Phys. Rev. Lett. {\bf 107}, 244101 (2011).
\bibitem{Skardal2011PRE} P. S. Skardal, E. Ott, and J. G. Restrepo, Phys. Rev. E {\bf 84}, 036208 (2011).
\bibitem{Komarov2013PRL} M. Komarov and A. Pikovsky, Phys. Rev. Lett. {\bf 111}, 204101 (2013).
\bibitem{Erdos1960} P. Erd\H{o}s and A. R\'{e}nyi, Pub. Math. Inst. Hung. Acad. Sci. {\bf 5}, 17 (1960).
\bibitem{Bekessy1972} A. Bekessy, P. Bekessy, and J. Komlos, Stud. Sci. Math. Hung. {\bf 7}, 343 (1972).
\bibitem{Newman2003PRE} M. E. J. Newman, Phys. Rev. E {\bf 67}, 026126 (2003).
\end{thebibliography}

\end{document}